\magnification=1200
\baselineskip=18pt plus 1pt minus 1pt
\tolerance=1000
\parskip=10pt
\parindent=25pt
\nopagenumbers
\font\rBig=cmr10 scaled\magstep 3
\font\rbig=cmr10 scaled\magstep 1
%\input phyzzx
%\pubnum{04-90}

\def\pb{$p$-brane}
\def\pbs{$p$-branes}
\def\vlambda{{\vec \lambda}}
\def\vzeta{{\vec \zeta}}

\def\ds{ D \kern-.63em / }
\def\ref#1{$\rm^{#1}$}

\baselineskip=10pt
%\pubtype={}
\rightline {Aug, 1994}
\vskip 4 truecm
\centerline{{\rBig Sphere-like Solutions in Surface Functional Theory}}
\centerline{{\rBig and Dirac's Membrane Model}}
\vskip 1truein
\centerline{{\bf Choon-Lin Ho}}
\medskip
\centerline{{\sl Department of Physics,  Tamkang University,
Tamsui, Taiwan 25137,  R.O.C.}}

\baselineskip=18pt plus 1pt minus 1pt

\vskip 1 truein

\centerline{{\rbig  Abstract}}
A surface functional theory for $p$-dimensional extended objects, the \pbs\ ,
was proposed in previous papers.  The field equations for toroidal \pbs\ was
exactly solved in $d=p+2$ dimensions, yielding equally spaced mass-squared
spectrum with massless states.  In this paper, we obtain the asymptotic
distribution of mass spectrum in the point-particle limit of the
theory with sphere-like membranes
($p=2$) in $d=4$ dimensions.  Similarity between this spectrum and that
obtained in the Dirac's membrane model of electron is discussed.
\vfill\eject

\pageno=2
\footline={\hss\tenrm\folio\hss}

\leftline{\bf 1. Introduction}

The revival of interest and the excitement aroused by string theory has
stimulated interest in the theory of other higher dimensional extended objects,
generally known as \pbs\ [1].  Green-Schwartz-type actions had been
constructed for \pbs\ , which are closely related to the four classical
superstring theories [2].  The most interesting Green-Schwartz-type super \pb\
theories is the 11D supermembrane theory, since it could have maximally
extended supergravity as its low energy approximation.  It is also realized
that membrane theory has very interesting algebraic structure : it has an
area-preserving algebra which contains the Virasoro algebra as a subalgebra
[3]. This area-preserving algebra is also closely related to a new field in
mathemetics, namely, the quantum group [4].  In view of the beautiful and rich
structures inherent in \pb\ theory, a better understanding of the theory is
desirable: it might provide new insight to string theory and related
mathematics.

On the other hand, it is tremendously difficult to further develop \pb\ theory
as unified theories.  Firstly, it is very difficult to quantize \pbs, because
the \pb\ actions are inextricably nonlinear.  Most quantization schemes
are only semiclassical.  Secondly,  one has to settle the question of the
existence of massless states and the discreteness of the mass spectrum in \pb\
theory.  It has been argue that the 11D supermembrane theory is not likely
to contain a massless state, and that its spectrum is continuous [5].

In [6], a surface functional theory of geometric \pbs\ was proposed, which
keeps the fundamental reparametrization invariance manifest.  There
both strings (regarded as $1$-branes) and general \pbs\ were given a unified
treatment.  The \pb\ fields are considered as multicomponent surface
functionals.  Dirac-type field equations were obtained for the surface
functionals.  Generalized Dirac algebras were introduce in order to
linearize the Dirac-Nambu-Goto actions of \pbs. In this way quantization of
\pbs\
could be carried out in a completely quantum-mechanical manner.  The field
equations were solved exactly for toroidal \pbs\ in $d=p+2$ dimensions and
$X_0=\tau$ gauge,  yielding an equally spaced mass-squared
(string-like) spectrum.  The ground states are always massless.  It was later
shown that the above results are the manifestation of $N=p+1$ hidden
supersymmetries in the theory [7].

Of course, solutions with toroidal \pbs\ represent only a very small fraction of all
the possible solutions that our theory admits.  It is, however, very
difficult to find exact solutions with other topological configurations.

In this paper, we shall attempt to find the asymptotic distribution of mass
spectrum in the point-particle limit of our theory with sphere-like membranes in four dimensions.  The
surface functional theory is first reviewed in Sect. 2.  The mass spectrum of
the theory with sphere-like membranes is then obtained in Sect. 3.   In Sect.
4, we discuss similarity between our mass spectrum and that obtained in
Dirac's membrane model [8], initially proposed to
explain the existence of muon.   Sect. 5 concludes the paper.

\vskip 1 truecm

\leftline{\bf 2. Surface Functional Theory for $p$-branes}

We first briefly review the essence of the surface functional theory of
$p$-branes as proposed in Ref. [6].

Consider a \pb\ moving in a $d$-dimensional flat Minkowski space-time.  The
classical Dirac-Nambu-Goto action for \pb\  is given by the volume
of the world volume swept out by the extended object in the course of its
evolution from some initial to some final configuration:
$$\eqalign{
&S =-{1\over \kappa} \int d^{p+1} \xi~
     \sqrt{(-1)^p ~det~{\hat h}_{\alpha\beta}}~,  \cr
&{\hat h}_{\alpha\beta}(\xi)=\partial_\alpha X^\mu ~\partial_\beta X_\mu~,
 \cr } \eqno(2.1) $$
where $X^\mu(\xi)~(\mu =0,\ldots,d-1)$ and
$\xi^\alpha =(\tau,\sigma_1,\ldots,\sigma_p)~(\alpha=0,\ldots,p)$ are
space-time and world volume coordinates, respectively. $\kappa$ is a
proportional constant the inverse of which could be thought of as the tension
of the \pb .  Space-time metric is taken to be
$\eta^{\mu\nu}={\rm diag} (1,-1,\ldots,-1)$.
One could then derive a set of primary constraints from this action.  It is
found that the
Poisson-bracket algebra of these constraints closes only weakly, except in the
case for string ($p=1$).  Furthermore, when the theory is quantized, the
commutator algebra of the constraints would give rise to operator anomalies.

These difficulties are avoided in the surface functional theory
proposed in [6].  In this approach, the basic entities are
taken to be
$p$-dimensional surfaces, or $p$-surfaces.  Fields are functions of
$p$-surfaces, and hence are called $p$-surface functionals.  These surface
functionals depend only on the location and topology of the $p$-surfaces, but
not
on their parametrization.  In this way, the important symmetry inherent in \pb\
dynamics, namely the reparametrization invariance, is maintained manifestly in
the theory.  The equations of motion for the surface functionals are required to
reproduce the Dirac-Nambu-Goto \pb\ dynamics in the $X_0=\tau$ gauge.  This
connection
is made in accordance with Dirac's treatment of spin-${1\over 2}$ particles.
This leads to the following covariant equations of
motion for multi-component fields $\Psi$ of space-time $p$-surfaces:
$$\eqalign{
&\int d^p\sigma\left[\Gamma^\mu(\sigma)p_\mu(\sigma)-
{\sqrt{(-1)^p~{\hat h}}\over \kappa}~\Lambda (\sigma)\right]
\Psi[X^\mu(\sigma)]=0~,\cr
&\left(\partial_kX^\mu\right)p_\mu\Psi=0~,\cr
&k=1,\ldots,p,\qquad \sigma\equiv \{\sigma_1,\ldots,\sigma_p\}~.\cr}
\eqno(2.2)$$
Here $p_\mu=i{\delta\over \delta X^\mu(\sigma)}$, and
${\hat h} = {\rm det}~{\hat h_{jk}}\ (j,k=1,\ldots,p)$.
$\Gamma^\mu(\sigma)$ and $\Lambda (\sigma)$ are generalised Dirac matrices.
If we define a totally anti-symmetric tangent
tensor
$$t_{\mu_1\cdots\mu_p}\equiv {1\over \sqrt{(-1)^p~{\hat h}}}
~{\partial\left(X_{\mu_1},\cdots,X_{\mu_p}\right)\over
\partial\left(\sigma_1,\cdots,\sigma_p\right)},\eqno(2.3)$$
with $t^2=(-1)^pp!$,  then $\Gamma^\mu(\sigma)$ and $\Lambda (\sigma)$
are given by
$$\eqalign{
\Gamma^\mu(\sigma)&=a_1\Gamma^\mu
         +{a_2\over p!}~\Gamma^{\mu\nu_1\cdots\nu_p}_A~t_{\nu_1\cdots\nu_p}
         +{a_3\over p!}~\Gamma^{\mu\nu_1\cdots\nu_p}_M~t_{\nu_1\cdots\nu_p} \cr
\Lambda (\sigma)   &=1 . \cr} \eqno(2.4)$$
[Actually $\Lambda (\sigma)$ can contain arbitrary powers of
$t_{\nu_1\cdots\nu_p}$,  we set $\Lambda (\sigma)=1$ for simplicity].
Here $a_1,a_2$ and $a_3$ are arbitrary constants.  $\Gamma^\mu$ are the usual
Dirac matrices.  $\Gamma^{\mu_1\cdots\mu_{p+1}}_A$ are totally anti-symmetric
in their indices, while
$\Gamma^{\mu_1\cdots\mu_{p+1}}_M$ are only anti-symmetric in the last $p$
indices and are taken to be traceless, in view of the second equation in (2.2).
Algebra satisfied by the $\Gamma_M$ -piece is generally very complicated, and
equation which involves the $\Gamma_M$-piece is also difficult to solve in
general.  In this paper, we shall consider only the $\Gamma^\mu$ and the
totally anti-symmetric $\Gamma^{\mu_1\cdots\mu_{p+1}}_A$ piece.  Full equations
containing the $\Gamma_M$ piece for membranes and strings are given in [6].

Eq. (2.2) can now be written as
$$\left[a_1\Gamma^\mu P_\mu+{a_2\over p!}\Gamma^{\mu_1\cdots\mu_{p+1}}_A
P_{\mu_1\cdots\mu_{p+1}}-{V_p\over \kappa}\right]\Psi=0~, \eqno(2.5)$$
where
$$\eqalign{
&V_p=\int d^p\sigma\sqrt{(-1)^p~{\hat h}}~,\cr
&P_\mu =\int d^p\sigma p_\mu~,\cr
&P_{\mu_1\cdots\mu_{p+1}}={1\over p+1}\int d^p\sigma
\left[\sum_\rho{\rm
sign}(\rho)~t_{\mu_1\cdots\mu_p}p_{\mu_{p+1}}\right]~.\cr}$$
Here $\rho$ represents cyclic permutation of the indices
$\mu_1$,\dots,~$\mu_{p+1}$.  The two constants $a_1$ and $a_2$  satisfy
$${a_1}^2+(-1)^p{a_2}^2=1~. \eqno(2.6)$$
The $\Gamma$'s matrices obey the following anti-commutation relations:
$$\eqalignno{
\{\Gamma^\mu,\Gamma^\nu\}&=2{\eta}^{\mu\nu}~,&(2.7)\cr
\{\Gamma^{\mu_1\cdots\mu_{p+1}}_A,\Gamma^{\nu_1\cdots\nu_{p+1}}_A\}
&=2\eta^{\mu_1\cdots\mu_{p+1},\nu_1\cdots\nu_{p+1}}~,&(2.8)\cr
{\rm All\ \ others}&=0~. \cr}$$
The generalized $\eta$-tensor is defined by
$$\eta^{\mu_1\cdots\mu_{p+1},\nu_1\cdots\nu_{p+1}}
\equiv\left|\matrix{
\eta^{\mu_1\nu_1}&\eta^{\mu_1\nu_2}&\cdots&\eta^{\mu_1\nu_{p+1}}\cr
\eta^{\mu_2\nu_1}&\eta^{\mu_2\nu_2}&\cdots&\eta^{\mu_2\nu_{p+1}}\cr
\vdots&\vdots&\ddots&\vdots\cr
\eta^{\mu_{p+1}\nu_1}&\eta^{\mu_{p+1}\nu_2}&\cdots&\eta^{\mu_{p+1}\nu_{p+1}}
\cr}\right|~. \eqno(2.9)$$
The algebras (2.7) and (2.8) are Clifford algebras of dimensions $d$ and
${d(d-1)\cdots(d-p)\over (p+1)!}$ respectively.
As such, the field
functional $\Psi$ transforms non-trivially under Lorentz transformations.

The equation of motion can be greatly simplified
in $d=p+2$ dimensions and $X_0=\tau$ gauge.  Under these conditions, the
$p$-surfaces become hypersurfaces in $(p+1)$-dimensional space.  Eq.(2.5)
now reduces to
$$\left[a_1\Gamma^\mu P_\mu + ia_2\Lambda^0~Q
 - {V_p\over \kappa}\right]\Psi=0~. \eqno(2.10)$$
The dual matrices
$$\Lambda^\mu\equiv {1\over (p+1)!}\epsilon^{\mu\mu_1\ldots\mu_{p+1}}
~\Gamma_{\mu_1\ldots\mu_{p+1}} \eqno(2.11)$$
satisfy
$$\{\Lambda^\mu,\Lambda^\nu\}=(-1)^{p+1} 2 \eta^{\mu\nu}~, \eqno(2.12)$$
which is, up to a sign, just the usual Dirac algebra.
Only $\Lambda^0$ contributes in (2.10).  The operator $Q$ is given by
$$Q = \int d^p\sigma~{\bf N}(\sigma)\cdot{\delta\over
\delta{\bf X}(\sigma)}~. \eqno(2.13)$$
${\bf N}(\sigma)$ is the unit normal vector at each point of the $p$-surface
in $E^{p+1}$, and is related to the tangent tensor by
$$N_j = {1\over p!}\epsilon_{i_1\ldots i_p j}
~t_{i_1\ldots i_p}~, \quad \quad  j,i_k=1,2,\ldots,p+1~.\eqno(2.14)$$
The operator $Q$ is in fact the generator of small
deformation of $p$-surface along the normal direction at each of its points,
and is therefore related to the
differential-geometric properties of $p$-dimensional hypersurfaces, as will be
shown below.

Let
$h={\rm det}~ h_{ij}={\rm det}~ (\partial_i {\bf X}\cdot \partial_j {\bf X})$
and $H_r$ be the $r$-th mean curvatures of the hypersurface.
Then
we can define the following $\sigma$-reparametrization invariant coordintes
$$\eqalignno{
h_\ell &= \int d^p\sigma \sqrt{h}~H_\ell~, \qquad \qquad \ell=0,1,\ldots,p~,\cr
{\bf y}& ={1\over h_0} \int d^p\sigma \sqrt{h}~{\bf X}~, \cr
\vlambda_r &= {p\choose r} \int d^p\sigma \sqrt{h}~ ({\bf X-y})~H_r~, \cr
\noalign{\hbox{\rm and}}
\vzeta_r  &= {p\choose r} \int d^p\sigma \sqrt{h}~{\bf N}~H_r~,
\quad r=1,\ldots,p~. &(2.15)\cr}$$
We note here that $h_0=V_p$, and $h_p$ is the total Gauss curvature.  {\bf y}
is the center-of-mass coordinates of the $p$-surface.
In terms of the coordinates $\{\tau$, $h_\ell$, $\bf y$, $\vlambda_r$,
$\vzeta_r\}$, we have
$$P^\mu =\left(i{\partial\over \partial\tau}, -i{\partial\over
\partial{\bf y}}\right)~,\eqno(2.16)$$
and
$$\eqalignno{
Q&=-\left[ph_1{\partial\over \partial h_0}+(p-1)h_2{\partial\over \partial h_1}
+\cdots + h_p{\partial\over \partial h_{p-1}}\right]+{\vlambda_1\over h_0}\cdot
{\partial\over \partial{\bf y}} \cr
&\phantom{=}\;+\left[-2\vlambda_2+\vzeta_1+p{\vlambda_1\over h_0}h_1\right]\cdot
{\partial\over \partial\vlambda_1} \cr
&\phantom{=}\;+\left[-3\vlambda_3+\vzeta_2+{p\choose 2}{\vlambda_1\over h_0}h_2
\right]\cdot{\partial\over \partial\vlambda_2}~-2\vzeta_2\cdot{\partial\over
\partial\vzeta_1} \cr
&{\qquad \qquad \qquad \qquad \vdots} \cr
&\phantom{=}\;+\left[-(r+1)\vlambda_{r+1}+\vzeta_r+{p\choose r}
{\vlambda_1\over h_0}h_r\right]\cdot{\partial\over \partial\vlambda_r}
~-r\vzeta_r\cdot{\partial\over \partial\vzeta_{r-1}} \cr
&{\qquad \qquad \qquad \qquad \vdots} \cr
&\phantom{=}\;+\left[\vzeta_p+{\vlambda_1\over h_0}h_p\right]\cdot
{\partial\over \partial\vlambda_p}~-p\vzeta_p\cdot{\partial\over
\partial\vzeta_{p-1}} . &(2.17)\cr}$$
It can be easily checked that the total Gauss curvature, $h_p$ , and
the coordinate $\zeta_p$ are constants of motion.

\vskip 1truecm

\leftline{\bf 3. Sphere-like Solutions in the Surface functional Theory for
Membrane}

We now proceed to solve the field equation (2.10) in this section.  In
general, the field equation (2.10) together with equations (2.16) and (2.17)
is still difficult to solve.  In [6] the field equation is exactly solved in
$p+2$ dimensions for toroidal $p$-branes
for which $h_2 = h_3 = \ldots = h_p = 0$, and $h_1\not= 0$,
yielding the string-like mass-squared spectrum with massless states.
But for $p$-branes of other topological configurations, the field equations are
difficult to solve.  Here we shall consider the asymptotic solutions for
sphere-like membranes in $d=4$ with constant $h_2 >0$. Note that $h_2$ is a
constant of motion, since $[Q,h_2]=0$.

We shall look for positive energy solutions which have the following form
$$\Psi =e^{-ip\cdot y}U({\bf p, y}, h_2 ,\vlambda_r ,\vzeta_r)\Phi
(h_0,h_1,h_2). \eqno(3.1)$$
with $y_0\equiv \tau$.  We require that $U$ satisfies
$$Q (e^{-ip\cdot y}U) = 0 \eqno(3.2)$$
If this is so, then one can factor out the factor $(e^{-ip\cdot y}U)$ from
the field equation.  It is not too difficult to find the required form of $U$
that solves eq.(3.2).  Let us note that in the case we are interested in, the
operator $Q$ has the following form,
$$\eqalignno{
Q&=-2h_1{\partial\over \partial h_0}-h_2{\partial\over \partial h_1}
+{\vlambda_1\over h_0}\cdot {\partial\over \partial{\bf y}} \cr
&\phantom{=}\;+\left[-2\vlambda_2+\vzeta_1+2{\vlambda_1\over
h_0}h_1\right]\cdot {\partial\over \partial\vlambda_1} \cr
&\phantom{=}\;+\left[\vzeta_2+{\vlambda_1\over
h_0}h_2 \right]\cdot
{\partial\over \partial\vlambda_2}~-2\vzeta_2\cdot{\partial\over
\partial\vzeta_1}. &(3.3)\cr}$$
The required $U$ is found to be:
$$U=\exp\biggl\{-{i\over h_2}{\bf p}\cdot [{\vlambda}_2 +
{\vzeta}_1/2 ]\biggr\}.\eqno(3.4)$$
With (2.16), (3.1), (3.2) and (3.3), the field equation (2.10) reads
$$\left[a_1\Gamma^\mu p_\mu + i  a_2\Lambda^0 Q
- {h_0\over \kappa}\right]\Phi (h_0,h_1,h_2 ) = 0~, \eqno(3.5)$$
where $p_\mu$ is the energy-momentum vector, and the operator $Q$ is now given
by
$$Q=-2h_1{\partial\over \partial h_0}-h_2{\partial\over \partial h_1}~.
\eqno(3.6)$$
To solve eq.(3.5), we take the
explicit representations of the $\Gamma^\mu$ and $\Lambda^\mu$ matrices as
follows:
$$\eqalign{
\Gamma^0 &=\sigma_1,\Gamma^1 =i\sigma_2,\Gamma^2 =i\sigma_3\sigma_1,
\Gamma^3 =i\sigma_3\sigma_2,\quad \ldots\cr
{\bar \Gamma} &=\sigma_3\sigma_3,\qquad
\Lambda^\mu = i{\bar \Gamma}\otimes\Gamma^\mu ~.\cr} \eqno(3.7)$$
\smallskip
Here the constants $a_1$ and $a_2$ are taken to be real (and positive,
without loss of generality) so as to ensure that the
mass squared, $p^2=m^2$, is bounded from below.  The wavefunction $\Phi$ has
$16$ components, which can be decomposed into four
$4$-component fields $\Phi_{ab}$.  The indices $a$ and $b$
( = $+$ or $-$) denote eigenvalues of ${\bar \Gamma}$ and
$I\otimes{\bar \Gamma}$
where $I$ is a $4\times 4$ matrix of unity.  After taking the square of
eq.(3.5) and rearranging the components, we obtain
$$\eqalign{
{a_1}^2 m^2 (\Phi_{++}\pm \Phi_{+-})
&=\left[-a_2^2 Q^2 +{h_0^2\over \kappa^2}
\pm {2h_1\over \kappa}a_2\right] (\Phi_{++}\pm \Phi_{+-})~, \cr
{a_1}^2 m^2 (\Phi_{-+}\pm \Phi_{--})
&=\left[-a_2^2 Q^2 +{h_0^2\over \kappa^2}
\mp {2h_1\over \kappa}a_2\right] (\Phi_{-+}\pm \Phi_{--}) ~. \cr} \eqno(3.8)$$

This equation is still difficult to solve.  Let us then consider only the
high membrane tension limit (also called the ``point particle limit") in which
$\kappa\rightarrow 0$.  In this case we can neglect terms in the first powers
of $1/\kappa$ in eq.(3.8).  The original eigenvalue problem now reduces to
a scalar problem:
$$
{a_1}^2 m^2 \Phi
=\left[-a_2^2 Q^2 +{h_0^2\over \kappa^2}\right] \Phi~, \eqno(3.9)$$
with the boundary condition $\Phi \rightarrow 0$ as $h_0, h_1\rightarrow
\infty$.   With the expression of $Q$ in (3.6), eq.(3.9) becomes:
$$\left(4h_1^2 {\partial^2\over \partial h_0^2}
+4h_1h_2 {\partial^2\over \partial h_0\partial h_1}
+h_2^2 {\partial^2\over \partial h_1^2}
+2h_2 {\partial\over \partial h_0}
-{1\over a_2^2\kappa^2} h_0^2 + {a_1^2 m^2\over a_2^2}
\right)\Phi =0. \eqno(3.10)$$
If we denote the coefficients of the first three derivative operators in (3.10)
by $c_1, 2c_2$ and $c_3$ respectively, then the discriminant of (3.10) [9]
is given by
$$\eqalign{
\Delta &\equiv c_2^2-c_1c_3\cr
       &=  (2h_1h_2)^2 - (4h_1^2)(h_2^2)\cr
       &= 0.\cr} \eqno(3.11)$$
This indicates that eq.(3.10) is of the parabolic type.
That means solutions of
the differential equation can be classified by family of curves along
which the basic variables satisfy the equation
$${dh_1\over dh_0} = {c_2\over c_1}={h_2\over 2h_1}~, \eqno(3.12)$$
or $h_1^2 - h_0 h_2 ={\rm constant}$.
Such curves are called the characteristics of the differential equation.
Hence of the two variables $h_0$ and $h_1$ only one is independent.
We can therefore transform the partial differential equation (3.10) into an
ordinary one (the canonical form)  by  the following change of variables:
$$\eqalign{
\xi(h_0,h_1) &= h_0\qquad,\cr
\eta(h_0,h_1)&= h_1^2 - h_0 h_2~.\cr}\eqno(3.13)$$
In terms of these new variables, eq.(3.10) becomes
$$4(\eta+h_2\xi){d^2\Phi\over d\xi^2}
+ 2h_2{d\Phi\over d\xi}
+ \left({a_1^2 m^2\over a_2^2} - {1\over a_2^2\kappa^2}\xi^2\right)\Phi=0~.
\eqno(3.14)$$
Note that $\eta$ is a constant of motion as $[Q,\eta]=0$.  Hence the evolution
of the system is along the characteristic curve corresponding to fixed $\eta$.

We now solve (3.14) for sphere-like membranes.  For a regular sphere of
radius $r$, one has $h_0=4\pi r^2$, $h_1=4\pi r$, and $h_2=4\pi$ from the
Gauss-Bonnet theorem, and hence $\eta=0$.  We consider membrane
configurations corresponding to $\eta=0$ as sphere-like in our theory.
Restricting ourselves to sphere-like membranes, eq.(3.14) reads
$$4\xi{d^2\Phi\over d\xi^2}
+ 2{d\Phi\over d\xi}
+ \left({a_1^2 m^2\over a_2^2 h_2} - {1\over a_2^2\kappa^2 h_2}
\xi^2\right)\Phi=0~. \eqno(3.15)$$

To solve for the eigenvalues of (3.15), we make a further change of variable.
Let $x\equiv ({a_2\kappa h_2^{1/2}})^{-{1\over 3}}\xi^{1\over 2}$.
Then (3.15) is changed into a Schr\"odinger-type equation:
$${d^2\Phi\over dx^2} + (\lambda^2 - x^4)\Phi =0~, \eqno(3.16)$$
with boundary condition $\Phi \rightarrow 0$ as $x\rightarrow \infty$.
Here the eigenvalues $\lambda$ is defined as
$$\lambda\equiv {a_1\over a_2^{2/3}}\left({\kappa\over h_2}\right)^{1\over
3} m~.\eqno(3.17)$$
The asymptotic  distribution of the eigenvalues of eq.(3.16) has been well
studied in the literature [10].  Applying the WKBJ method, one gets
$$\lambda_n =\left[ {(n+{1\over 2})\pi\over 2\int_0^1 \sqrt{1-t^4} dt}
\right]^{2\over 3}~.\eqno(3.18)$$
for large positive integers $n$.
The integral in (3.18) can be expressed in terms of the $\Gamma$ functions:
$$\int_0^1 \sqrt{1-t^4} dt ={\Gamma({1\over 4})  \Gamma({3\over 2})
\over 4\Gamma({7\over 4})}~.\eqno(3.19)$$
From (3.17) and (3.18) one gets the mass spectrum of our theory:
$$m=\left[{a_2^2 h_2\over 4 \kappa a_1^3}\right]^{1\over 3}
\left[ {(n+{1\over 2})\pi\over \int_0^1 \sqrt{1-t^4} dt}\right]^{2\over
3}~.\eqno(3.20)$$
Hence for sphere-like membranes in our theory, the mass vaires as $m \sim
n^{2\over 3}$ for sufficiently large
$n$. For comparision, let us recall that for toroidal membranes, one has
the string-like spectrum
$m\sim n^{1\over 2}$ [6].

The spectrum (3.20) turns out to be very similar to that obtained by Dirac in
a model of electron proposed in 1962 [8].  We shall now turn to discussing the
connection between the two spectra.

\vskip 1 truecm

\leftline{\bf 4. Dirac's Membrane Model}

In an attempt to explain the existence of muon, which has properties so
similar to the electron except the mass , Dirac proposed a membrane model of
the electron [8].   There he considered the classical electron as a charged
conducting surface, with a surface tension to prevent it from flying apart
under the repulsive forces of the charge.  The action of the system thus
consists of two parts:  the Maxwell term outside the surface, and a surface
tension term on the surface.  The second part is indeed the action of membrane.
Such an electron has an equilibrium state with spherical symmetry (radius
$\rho$), and if
disturbed its shape and size oscillate.  Dirac deduced the equation of motion
of the extended object from an action principle and a Hamiltonian formulation.
From the classical equation of motion he showed that the equilibrium radius
$\rho=a$ of the electron is given by
$$a^3 = {e^2\over 4 \omega}~, \eqno(4.1)$$
where Dirac's membrane tension $\omega$ is related to our $\kappa$ by $\omega
\equiv {4\pi \over
\kappa}$.  To connect $a$ with the mass $m_e$ of the electron, he considered
the total energy of an electron instantaneously at rest.  The total energy
$E$ consist of two parts:  the electrostatic energy of the Couloumb
field, $e^2/2\rho$, and a surface tension energy proportional to $\rho^2$:
$$E={ e^2\over 2 \rho}  + \beta \rho^2~. \eqno(4.2)$$
As the minimun value of $E$ must occur when $\rho=a$, one obtained
$$\beta ={ e^2\over 4 a^3}~.\eqno (4.3)$$
This shows that $\beta=\omega$, and that the surface energy in the equilibrium
state is half the electrostatic energy.  Hence the total energy is $E=3e^2/4a$.
But since $E=m_e (c=1)$, one gets
$$a={3e^2\over 4m_e}~.\eqno(4.4)$$
To find the energy or mass spectrum of the excited states, Dirac
proceeded to formulate a quantum theory of the model, hoping that the first
excited state would give the mass of the muon.  For this purpose Dirac had
to base his formulation on his then newly developed method of constraint
quantisation [11].  For spherically symmetric motion,  he showed that the only
independent dynamical variables are the radius $\rho$ of the electron and its
conjugate momentum $P$.  The corresponding Hamiltonian of the symmetric
system was then found to be
$$H = (P^2 + \omega^2 \rho^4)^{1\over 2} + e^2/2\rho~. \eqno(4.5)$$
Note that (4.5) reduces to (4.2) when $P=0$.
The Schr\"odinger equation with this Hamiltonian is difficult to solve, and
Dirac obtained the mass spectrum by means of the Bohr-Sommerfeld
quantisation rule :
$$2\pi n = 2\int^{\rm max}_{\rm min} P d\rho~. \eqno(4.6)$$
From (4.5) we have
$$P^2 = \left(E-{e^2\over 2\rho}\right)^2 - \omega^2\rho^4~.\eqno(4.7)$$
Let us define $x=\rho/a$ and $l=4aE/e^2$.  Then eq.(4.6) becomes
$${4n\pi\over e^2}=\int\left(\left(l-{2\over x}\right)^2-x^4\right)^{1\over 2}
dx~.\eqno(4.8)$$
For large values of $E$,  eq.(4.8) is approximately given by
$$\eqalign{
{4n\pi\over e^2}&=\int_0^{l^{1\over 2}}\left(l^2-x^4\right)^{1\over 2} dx \cr
              &=l^{3\over 2} \int_0^1 \sqrt{1-t^4} dt~.\cr}\eqno(4.9)$$
Eq.(4.9), together with eq.(4.4) and the definition of $l$, lead to the mass
spectrum in Dirac membrane theory:
$$m={m_e\over 3} \left({4\over e^2}\right)^{2\over 3}
\left[ {n\pi\over \int_0^1 \sqrt{1-t^4} dt}\right]^{2\over 3}~.
\eqno(4.10)$$
To Dirac's disappointment, the mass of the first excited state ($n=1$) turns
out to be $m=53 m_e$, which is only about a quarter of the observed muon mass
($m_\mu =210 m_e$).

Comparing eq.(4.10) and (3.20), we see that the mass spectra
in both
theories have the same asymptotic behaviour for large mass , {\it i.e.}, $m
\sim n^{2\over 3}$.
In fact, Dirac's mass spectrum could be obtained from the spectrum in our
theory by an appropriate choice of the membrane tension and the values of
$a_1$ and $a_2$.  To see this, let us assume that
the membrane tension is given by Dirac's expression (4.1) together with
eq.(4.4), then we get
$$\eqalign{
{1\over \kappa} &= {\omega\over 4\pi} \cr
                &= {4 m_e^3\over 3^3 e^4\pi}~.\cr}\eqno(4.11)$$
Substituing this expression into (3.20), and using the fact that $h_2=4\pi$,
we obtain
$$m={m_e\over 3} \left({4\over e^2}\right)^{2\over 3}
\left[{a_2^2\over 4a_1^3}\right]^{1\over 3}
\left[ {(n+{1\over 2})\pi\over \int_0^1 \sqrt{1-t^4} dt}\right]^{2\over 3}~.
\eqno(4.12)$$
Hence (4.10) and (4.12) will be identical (except for the irrelevant difference
in $n$ and $n+1/2$ for large $n$), provided that
$${a_2^2\over 4a_1^3}={{1-a_1^2}\over 4a_1^3}=1~.\eqno(4.13)$$
Here we have used the relation $a_1^2+a_2^2=1$ (eq.(2.6)).
Eq.(4.13) is solved by $a_1=0.557$ and $a_2=0.831$.
It is also of interest to note that if the mass of the first excited state
($n=1$) in our theory were to correspond to the muon mass, then we must have
$a_1=0.206$ and $a_2=0.979$ .  However, the power law $m\sim n^{2\over 3}$
does not allow one to account for both the observed values of the massess
of the muon and the $\tau$-meson.

\vfill\eject

\leftline{\bf 5. Summary}

We have obtained the asymptotic distribution of mass spectrum in the
point-particle limit of the
surface functional theory for sphere-like membranes in four dimensions.
By a specific choice of the membrane tension and the parameters in our theory,
we reproduce the mass spectrum in Dirac's membrane theory of electron.

\vskip 1 truein

\centerline{\bf Acknowlegments}
\bigskip
This work is supported by the R.O.C. Grant NSC83-0208-M032-017.

\vfill\eject

\def\prl{{\sl Phys. Rev. Lett.}}
\def\pr{{\sl Phys. Rev.}}
\def\pl{{\sl Phys. Lett.}}
\def\np{{\sl Nucl. Phys.}}
\def\ap{{\sl Ann. Phys. (N.Y.)}}

\def\cqg{{\sl Class. Quant. Grav.}}

\def\jmp{{\sl J. Math. Phys.}}
\def\cmp{{\sl Comm. Math. Phys.}}
\def\prsl{{\sl Proc. Roy. Soc. (London)}}

\centerline{\rBig References}

\vskip 1.3 truecm

\item{1.} For reviews, see eg: M.J. Duff, \cqg\ {\bf 5} (1988) 189; E.
Bergshoeff, E. Sezgin, and P.K. Townsend, \ap\ {\bf 185} (1988) 330, and
references therein. For some recent developments in gauge theory of membranes,
see eg:
A. Aurilia and E. Spallucci, \cqg\ {\bf 10} (1993) 1217; Preprint UTS-DFT-92-5,
Mar 1992.
\item{2.} J. Hughes, J. Liu, and J. Polchinski, \pl\ {\bf B180} (1986)
370;
E. Bergshoeff, E. Sezgin, and P.K. Townsend, {\it ibid.} {\bf B189} (1987) 75;
A. Achucarro, J. Evans, P.K. Townsend, and D.L. Wiltshire, {\it ibid.} {\bf
B198} 441.
\item{3.} E.G. Floratos and J. Iliopoulos, \pl\ {\bf B201} (1988) 237;
I. Antoniadis, P. Ditsas, E.G. Floratos, and J. Iliopoulos, \np\ {\bf B300}
(1988) 549; I. Bars, C.N. Pope, and E. Sezgin, \pl\ {\bf B210} (1988) 85;
D.B. Fairlie, P. Fletcher, and C.K. Zachos, \pl\ {\bf B218} (1989) 203.
\item{4.} M. Golenishcheva-Kutuzova and D. Lebedev, \cmp\ {\bf 148} (1992) 403;
J. Hoppe, M. Olshanetsky, and S. Theisen, {\it ibid.} {\bf 155} (1993) 429.
\item{5.} B. de Wit, J. Hoppe, and H. Nicolai, \np\ {\bf B305} (1988) 545;
B. de Wit, M. L\"uscher, and H. Nicolai, {\it ibid.} (1989) 135;
B. de Wit and H. Nicolai, in {\it Supermembranes and Physics in $2+1$
Dimensions:  Proceedings of the Trieste Conference}, World Scientific,
Singapore, 1990.
\item{6.} L. Carson and Y. Hosotani, \prl\ {\bf 56} (1986) 2144; \pr\ {\bf D37}
(1988) 1492; C.-L. Ho and Y. Hosotani, \prl\ {\bf 60} (1988) 885; C.-L. Ho,
\jmp\ {\bf 30} (1989) 2168; Ph.D. Thesis, Minnesota Report No.
UMN-TH-807/89,1989.
\item{7.} C.-L. Ho, \pr\ {\bf D42} (1990) 3439.
\item{8.} P.A.M. Dirac, \prsl\ {\bf A268} (1962) 57.
\item{9.} A. Sommerfeld, {\it Partial Differential Eqautions In Physics},
Academic Press, London, 1949.
\item{10.} E.C. Titchmarsh, {\it Eigenfunction Expansions Associated With
Second-Order Differential Equations}, Vol. 1, Clarendon, Oxford, 1962.
\item{11.} P.A.M. Dirac, \prsl\ {\bf A246} (1958) 326.
\vfill\bye